\documentclass[prl,twocolumn]{revtex4}
\usepackage{graphicx}
\usepackage{amssymb}
\begin{document}

\title{
Microscopic model for the higher-order nonlinearity in optical filaments
}

\author{A. Teleki$^1$, E.~M. Wright$^2$, M. Kolesik$^2$}
\affiliation{
$^1$Department of Physics, Constantine the Philosopher University, Nitra, Slovakia\\
$^2$College of Optical Sciences, University of Arizona, Tucson AZ 85721
}

\begin{abstract}
Using an exactly soluble one-dimensional atomic model we explore the idea that the recently
observed high-order nonlinearity in optical filaments is due to virtual transitions involving
the continuum states.  We show that the model's behavior is qualitatively comparable with
the experimentally observed cross-over from self-focusing to de-focusing
at high intensities, and only occurs at intensities which result in significant ionization. Based
on these observations, we conjecture that this continuum electron nonlinear refraction
exhibits strong memory effects, and most importantly, the change of its sign
is effectively masked by the de-focusing due to free electrons.
\end{abstract}

\maketitle

Since the first experimental observation of long-distance propagation
and optical filamentation in high-power femtosecond light pulses
more than a decade ago~\cite{BraunOL95}, it has been accepted that the main nonlinear
effects controlling the phenomenon are the optical Kerr effect and
de-focusing due to the free electrons generated by high-intensity
ionization. However,  Loriot et al. recently presented an experimental measurement
of higher-order, intensity-dependent Kerr nonlinearity in optical
filaments~\cite{loriot_measurement_2009,loriot_measurement_2010},
which exhibits a cross-over from self-focusing to de-focusing
at high intensities.
This development was quickly followed by a theoretical work~\cite{Bejot2010} which
concluded that the long-standing theory of optical filamentation
in gases needs to be changed in radical ways. In particular, it was
proposed that the occurrence of free electrons is not necessary
for formation of femtosecond filaments. At time of this writing,
the experiment still awaits
an independent corroboration. Moreover, a microscopic explanation
of the higher-order nature of the nonlinear refraction beyond the usual Kerr effect has
not been offered so far.  Using full quantum simulations of an atom subjected to a short pulse Nurhuda and co-workers
have attempted to deduce the form of the higher-order nonlinear refraction (see e.g. \cite{nurhuda}).
However, these efforts are hampered by the difficulty of separating the various contributions to the total
nonlinear optical response.

In this paper we advance the idea that higher-order nonlinear refraction can arise from virtual
transitions from the ground atomic state to continuum states and back to the ground state.  We term this continuum
electron nonlinear refraction in contrast to the more usual bound electron nonlinear refraction that involves virtual
transitions from the ground state to bound states and back to the ground state~\cite{Boyd}.
The bound electron nonlinear refraction is usually calculated within third-order perturbation theory
and leads to the familiar Kerr nonlinearity for transparent dielectrics. In contrast, as we shall see,
evaluation of the continuum electron nonlinear refraction involves a non-perturbative calculation that
leads to a nonlinear optical response that can cross over from self-focusing to self-defocusing at high
intensities.

To elucidate the physics of continuum electron nonlinear refraction we employ an exactly soluble one-dimensional
atomic model where the electron-ion interaction is modeled using an attractive delta-function potential,
sometimes termed 'delta-Hydrogen'.  This model has the virtues that in the absence of any external fields
it has only one bound state, the ground state, plus the continuum states, and the model remains soluble
even in the the presence of an external field.  This allows us to isolate the continuum electron nonlinear
refraction since there is no bound electron nonlinear refraction in this model.  We remark that the
one-dimensional atomic model has been explored extensively by the mathematics \cite{bookCycon,bookAlbeverio}
and physics
communities~\cite{villalba_particle_2009,peng_application_2006,uncu_solutions_2005,dunne_simple_2004,alvarez_perturbation_2004}.
Previous works involving light-matter interactions include tunneling ionization~\cite{elberfeld_tunneling_1988}
in strong laser fields~\cite{Geltman1978,cavalcanti_decay_2003} and also the time dependence of the survival
probability of the decaying ground state~\cite{arrighini_ionization_1982}.

The occurrence of higher-order nonlinear refraction has clear wide ranging implications for the dynamics
of self-focusing collapse and the formation of filamentation in general, and so a goal of our study is to
pose and answer, at least partially, the following questions raised by the Loriot's experiment:

\begin{itemize}
\setlength{\itemsep}{0cm}%
\setlength{\parskip}{0cm}%

\item[A)] What is the microscopic mechanism behind the intensity dependent nonlinearity?

\item[B)] Which occurs earlier, ionization or the higher-order Kerr focusing-defocusing cross-over?
\item[C)] To what extent is the higher-order nonlinearity instantaneous?
\item[D)] Perhaps most importantly, why no previous
experiments produced a clear evidence of the higher-order
nonlinearity?

\end{itemize}

In this work,
we address the above problems in the low-frequency limit, i.e. when the ratio of the
photon energy to the ionization potential is sufficiently small so that we can study the metastable
ground-state in a quasi-static approximation.
While it is obvious that the simplified one-dimensional atomic model employed here
is too simple to produce any quantitatively verifiable outputs, we contend that it provides insight
into the underlying physics of higher-order nonlinear refraction. This is supported by the fact that
higher-order nonlinear refraction appears to be universal, and in particular common to atoms and molecules,
and as such it should only depend of the most basic properties of the given system.

For the sake of reader's convenience, we next recall the definition
of the model, and its exact solution in the form of the Hamiltonian's
resolvent. Then we identify the relevant resolvent pole in the non-physical
sheet of the spectral parameter, and from it we calculate the field
dependent polarizibility of the ground state.

In a symbolic form the Hamiltonian can be written as is usual
in the physics literature, namely in terms of a delta-function potential
acting on a one-dimensional particle in a homogeneous external field:
\begin{equation}
H = \frac{-\hbar^2}{2 m} \frac{{\rm d}^2}{{\rm d}x^2} + \frac{-\hbar^2 A}{2 m} \delta(x) - e F x
\end{equation}
More precisely, the domain of the Hamiltonian consists of
locally absolutely continuous functions, $AC_{\rm loc}({\mathbb R})$,
with their derivative in $AC_{\rm loc}({\mathbb R\!\setminus\! 0})$,
which satisfy:
\begin{equation}
\frac{{\rm d}\psi(x)}{{\rm d}x}\bigg|_{x=0+} - \frac{{\rm d}\psi(x)}{{\rm d}x}\bigg|_{x=0-} = -A \psi(0)
\end{equation}
Here, the quantity $A$, which we assume to be positive, measures the strength of the delta-function potential.
The above condition ``encodes'' the delta-function in the domain of the Hamiltonian whose action is then defined
through the differential operator
\begin{equation}
H\psi(x) = \frac{-\hbar^2}{2 m}\frac{{\rm d}^2  \psi(x)}{{\rm d}x^2} - e F x\psi(x) \ ,
\label{eq:diff}
\end{equation}
which is the same as in the system with no contact potential.
It is also required that the result of (\ref{eq:diff}) is
quadratically integrable.
To simplify notation, we utilize
units in which $m=1/2, e=1$, and $\hbar = 1$.

The spectral properties of $H$ are
fully captured in its resolvent $(H-\lambda)^{-1}$ or, equivalently,
in the Green's function $G(x,x',\lambda) = \langle x|(H-\lambda)^{-1}|x'\rangle$
\begin{equation}
G(x,y,\lambda) = G_0(x,y,\lambda) + \frac{G_0(x,0,\lambda) G_0(0,y,\lambda)}{1/A - G_0(0,0,\lambda)} \ .
\label{eqn:GreenFull}
\end{equation}
where $G_0(x,x',\lambda)$ is the Green's function for $A=0$.
The above is an exact result and a consequence of the Krein's theorem~\cite{bookAlbeverio}.
We refer the reader to~\cite{cavalcanti_decay_2003} for an intuitive derivation.

With $A=0$,  $G_0(x,x',\lambda)$ satisfies the equation
\begin{equation}
-\partial_{xx}  G_0(x,y,\lambda) - (F x +\lambda) G_0(x,y,\lambda)  =  \delta(x-y) \ ,
\label{eqn:Green0}
\end{equation}
and can be expressed through a pair of solutions to the homogeneous equation:
\[
 G_0(x,y,\lambda) = \matrix{  {-\psi_L(x,\lambda) \psi_R(y,\lambda)}/{W(\psi_L,\psi_R)}  \  , \ \ \ x < y  \cr
                              {-\psi_R(x,\lambda) \psi_L(y,\lambda)}/{W(\psi_L,\psi_R)}  \  , \ \ \ x > y    }
\]
where the Wronskian $W(\psi_L,\psi_R) = \psi_L \partial_x \psi_R - \psi_R \partial_x \psi_L$,
and the left and right solutions $\psi_{L,R}$ must tend to zero at their respective
infinities
\[
\lim_{x\to -\infty} \psi_L(x,\lambda) = 0, \ \ \ \lim_{x\to +\infty} \psi_R(x,\lambda) = 0, \ \ {\rm Im}\{\lambda\}\ne 0 \ .
\]
In the upper half-plane of the spectral parameter, ${\rm Im}\{\lambda\}>0$, we can write  $\psi_{L,R}$ in terms of
Airy functions as follows:
\[
\psi_L(x,\lambda ) = {\rm Ai}(-\xi)
\ \ \
\psi_R(x,\lambda ) = {\rm Bi}(-\xi) + i {\rm Ai}(-\xi)
\]
with (assuming $F>0$)
\[
\xi = F^{1/3} (x + \lambda/F) \ \ \ {\rm and} \ \ \  W= -F^{1/3}/\pi
\]
The asymptotic expansions appropriate for the respective sectors of the complex plane~\cite{bookAbram} show
that at infinity these solutions behave as
\[
\psi_L \sim \exp{(-2 i \xi^{3/2}/3 )} \ \ \ {\rm and} \ \ \ \psi_R \sim \exp{(+2 i \xi^{3/2}/3 )}
\]
and therefore vanish exponentially for ${\rm Im}\{\lambda\}>0$  and
$x$ approaching negative and positive infinity, respectively.
Thus, $\psi_{L,R}$ is indeed the appropriate pair of solutions to calculate the Green's function. Equivalently,
they can be viewed as combinations of the Hankel functions ${\rm H}^{1,2}_{1/3}(2/3 \xi^{3/2})$.
This may be advantageous for ``traversing'' the entirety of the two spectral-parameter sheets since
all needed analytic continuations can be conveniently obtained with a single pair of functions.
However, we will restrict our attention to a sector in which the above Airy representations are
sufficient and easy to use.

\begin{figure}[t]
\vspace{3mm}
\centerline{
\scalebox{0.4}{\includegraphics{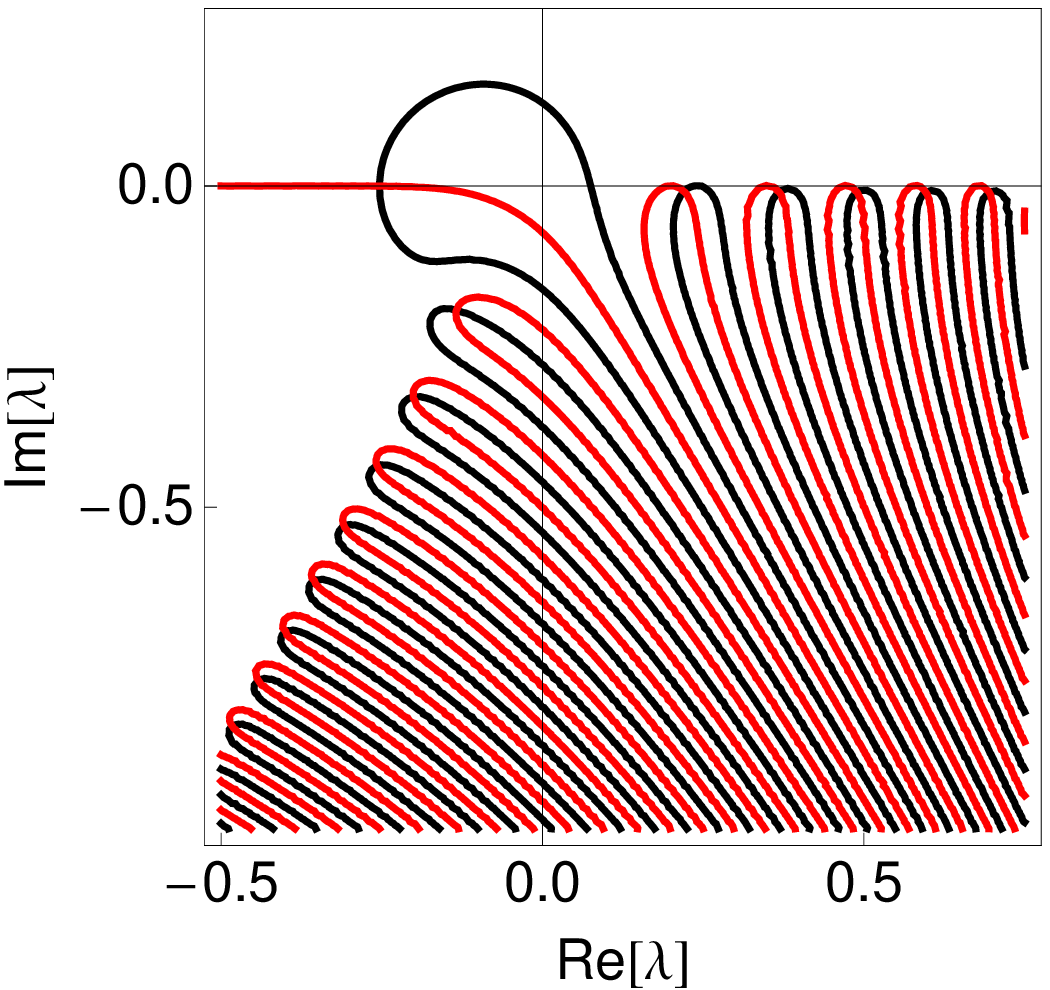}}
\scalebox{0.4}{\includegraphics{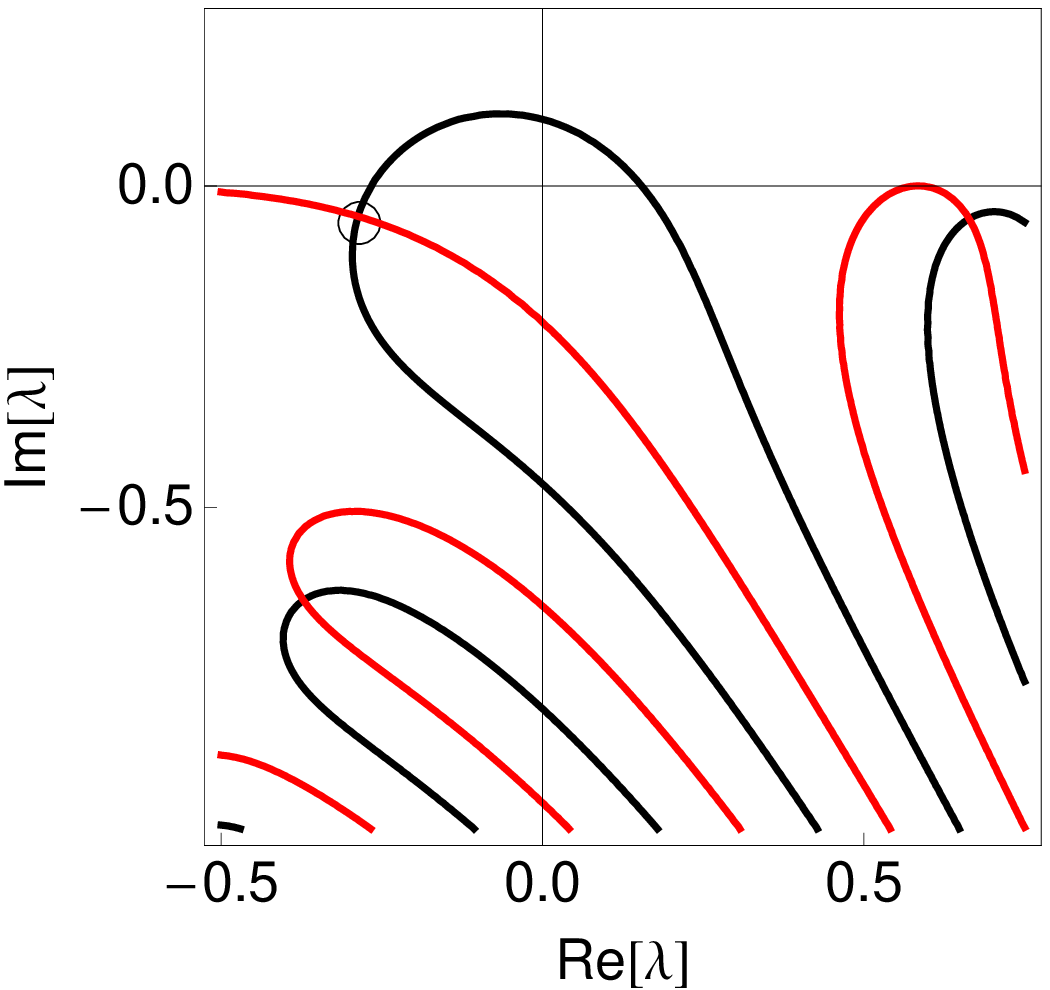}}
}
\caption{\label{fig:contours}
(Color online)
Contour plot of zeros of the real (black) and the imaginary (red)
parts of (\ref{eqn:pole}). Resolvent resonances appear where
the contours intersect. The row of solutions close to the real
axis correspond to the zeros of ${\rm Ai}$. Poles along
$e^{-2i\pi/3}$ are close to zeros of ${\rm Ci}^+$. The resonance
that originates in the bound state is marked by the circle in the
right panel. On the left, the field is weak, and the resonance location
is barely distinguishable from the ground-state energy $E_{\rm g}=-1/4$.
Left: $F=0.025$, right: $F=0.125$. $A=1$.
}
\end{figure}

Having obtained explicit expressions needed for the full resolvent (\ref{eqn:GreenFull}) we now turn
our attention to its poles. The denominator of the second part gives the equation to find the
resonances, namely $1 = A G_0(0,0,\lambda)$ which explicitly reads
\begin{equation}
1 = \frac{\pi A}{F^{1/3}} {\rm Ai}\bigg(\frac{-\lambda}{F^{2/3}}\bigg)
\left[ i {\rm Ai}\bigg(\frac{-\lambda}{F^{2/3}}\bigg) + {\rm Bi}\bigg(\frac{-\lambda}{F^{2/3}}\bigg) \right]
\label{eqn:pole}
\end{equation}
This equation has infinitely many solutions; for a small $F$, zeros of
${\rm Ai}$ and ${\rm Ci}^+ \equiv {\rm Bi} + i {\rm Ai}$ give rise to two families
of resonances. However, we are interested in another solution which converges, with $F\!\to\! 0$, to the
zero-field ground-state energy $E_{\rm g}=-A^2/4$ on the real axis while approaching from the lower half-plane.
Figure~\ref{fig:contours} illustrates all three types of solutions in weak and strong fields.
The resonance corresponding to the metastable ground state is marked by a circle.

The real and imaginary parts of this resonance pole location
$\lambda_{\rm R} = E_{\rm r} - i E_{\rm i}$ are related to the
ionization rate and nonlinear polarizibility, respectively.
Specifically, the ionization rate is proportional to its imaginary part
\[
w = \frac{1}{\tau} = \frac{2 E_i}{\hbar}  \ .
\]
A useful scaled quantity is the ionization rate in units of time given
by the oscillation period of the driving field. If we specify the
wavelength equivalently as the number $n$ of photon
energies needed to overcome the system's ionization potential $|E_{\rm g}|$,
then the corresponding scaled ionization rate reads
\[
w_n = 2 n\frac{E_i}{|E_{\rm g}|}
\]
This quantity tells us if we should expect an appreciable fraction
of atoms to be ionized within a single cycle of the optical field.
For argon with its ionization potential of 15.7~eV and for the wavelength
of 800~nm, $n=11$. We will therefore use $w_{11}$ to place our results in
the context of femtosecond filamentation.

To relate $E_{\rm r}$ to the nonlinear polarization, we express the ground-state energy shift
$E_{\rm r}-E_{\rm g}$ as the energy of an induced dipole in the electric field. In turn, the
induced dipole is expressed through the field-dependent polarizibility $\alpha(E)$:
\[
E_r-E_g = p E = \alpha(E) E^2
\]
The nonlinear component of the polarizibility, namely $\alpha_{\rm NL}(E) = \alpha(E) - \alpha(0)$
is the quantity we aim to compare to the results of the higher-order Kerr measurement.
As with the ionization rate, it is convenient to utilize a scaled form of
the nonlinear polarizibility, namely
\[
\chi_{\rm NL} = \alpha_{\rm NL}(E)/\alpha(0)
\]
This quantity, termed scaled nonlinear susceptibility,  measures the induced susceptibility
of refraction relative to the linear-regime susceptibility of the unperturbed system.

\begin{figure}[t]
\centerline{
\scalebox{0.45}{\includegraphics[clip]{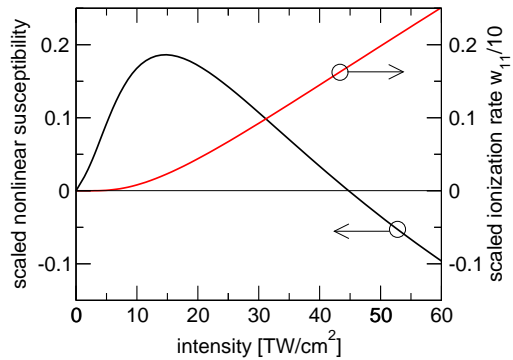}}
}
\caption{\label{fig:nlchi}
(Color online) Nonlinear susceptibility (black) and ionization
rate (red) obtained from the  exact solution for the resolvent pole.
}
\end{figure}

We next discuss the properties of the exact solution to the resonance
equation (\ref{eqn:pole}). We assume that the system has an ionization potential of
15.7~eV, i.e. that of argon, the incident pulse wavelength
is 800~nm, and we express the field strength $F$ equivalently through the
light intensity $I$.  Figure~\ref{fig:nlchi} depicts the result obtained from the
exact solution. The nonlinear susceptibility exhibits nearly linear increase
for low intensity, which corresponds to the usual Kerr nonlinearity when
the induced index of refraction is proportional to the light intensity.
At higher intensities, $\chi_{\rm NL}(I)$ saturates and subsequently
decreases into negative values. The zero crossing occurs at $I\approx 45$TW/cm$^2$.
For very high intensities (not shown) the scaled susceptibility approaches $-1$ which
corresponds to complete cancellation of the atom's linear susceptibility.
However, the extreme high-intensity regime is practically irrelevant
because the survival probability of the ground state vanishes rapidly.
The second curve in Fig.~\ref{fig:nlchi} shows the scaled ionization rate
for the 800~nm driving wavelength. We can infer that in the vicinity of the
susceptibility zero crossing, already a ten-cycle pulse is sufficient to reach ionization
probability of about eighty percent. We can also see that when the nonlinear
susceptibility starts to saturate, ionization already sets in.
In other words, the nontrivial behavior of the nonlinear susceptibility is
``synchronized'' with significant ionization.

For a complementary view of the interplay between ionization and polarizibility,
it is instructive to examine asymptotic solutions of (\ref{eqn:pole}) for weak fields.
Previous works~\cite{cavalcanti_decay_2003,Geltman1978} showed that the ionization rate is non-perturbative, and has the
functional form of the tunneling ionization. This means that there exist no Taylor
expansion in terms of the field strength. Because the polarizibility and ionization
are connected through the resolvent pole location, it is only natural to expect
that also the nonlinear index will be non-perturbative and lack proper Taylor
expansion; this is indeed the case as illustrated next.
In the lowest order (for $A=1$) the pole equation and its 
resonance solution $\lambda_{\rm R}$ become~\cite{cavalcanti_decay_2003}
\[
1 = \frac{i}{2 \sqrt{\lambda}} - \frac{e^{  -\frac{4  (-\lambda)^{\frac{3}{2}}}{3 F}  }}{4 \sqrt{\lambda}}
\ \ \ , \ \ \
\lambda_{\rm R} = -\frac{1}{4} - i e^{-\frac{1}{6F}}
\]
The imaginary part gives rise to the non-perturbative tunneling ionization rate~\cite{elberfeld_tunneling_1988}.
Here one can see that there are actually two small parameters, namely $F$ and $e^{-\frac{1}{6F}}$. We
develop the asymptotic solution of (\ref{eqn:pole}) in powers of these. For fixed expansion orders,
we obtain a converged approximation in a finite number of iterations.
The procedure shows that the real part of $\lambda_{\rm R}$ will also contain non-perturbative
terms, starting with  $e^{-\frac{1}{3F}}$. This means that neither
ionization nor the polarizibility can be developed as a Taylor expansion
in the field.  A second important point is that even very high-order asymptotic solution
can only reproduce the polarizibility curve roughly up to 10TW/cm$^2$ in Fig.~\ref{fig:nlchi}.
This is because the ``small'' parameter $e^{-\frac{1}{6F}}$ becomes large very quickly with
increasing field.
This sheds light on the fact that the measured
nonlinear coefficients seem to represent a ``divergent'' function.
The  representation in powers of intensity, $\sum n_{2k} I^k$, is actually ill suited
to represent this nonlinear behavior, and in our opinion contributes to the
rather large error bars.

Finally, let us summarize our finding from the exact solution of the model system,
and relate them to the questions posed in the introduction.

A) Most importantly, we can see that even a model with a single bound state exhibits
an intensity dependent nonlinearity very much similar to that observed by Loriot et al.,
and that it occurs in the comparable intensity range.
The absence of other bound states in the model indicates that the effect is due to virtual
transitions into continuum.
We speculate that, similarly to the high-harmonic generation,
the few necessary ``ingredients'' are limited to the existence of ground state and of
a continuum spectrum with the band-gap setting the scale for the intensity.

B) We showed that the nontrivial behavior in the polarizibility is intimately
connected with the ionization, which sets in as soon as the susceptibility starts
to saturate. Moreover, in this respect it is important to realize that in a real system
the excited bound states will give rise to the {\em additional} $n_2 I$ Kerr nonlinearity.
This will then shift the real part of the total susceptibility toward higher intensities,
while leaving the imaginary part intact. Consequently, the plasma generation will
start {\em before} the nonlinearity changes its sign to de-focusing. From here
we conjecture that in naturally occurring filaments, the higher-order
nonlinearity will be effectively masked by free electrons. We speculate that this may
very well be the answer to our question D), namely why previous experiments did not
noticed these effects.

C) The fact that it is the continuum states that give rise to the nontrivial
nonlinear behavior strongly suggests that the high-order nonlinearity will exhibit
finite memory whose influence is likely to become stronger with
increasing intensity. This is because unlike discrete states, the continuum of states
can form ``sufficiently many'' superpositions which can ``encode and remember'' the history of the system.
Of course, to verify this conjecture, time-resolved investigations will be necessary.
If corroborated, a departure from a truly instantaneous (or very fast) nonlinearity will have
an important consequence in decreasing the conversion efficiency in generation of
high frequencies and supercontinuum.

To conclude, while keeping in mind the simplicity of the studied model,
we believe that it gives us very important clues that will be eventually
useful for understanding microscopic mechanisms controlling filamentation processes
on few-femtosecond time scales.

\noindent
{\bf Acknowledgment:}
This work was supported by the FCVV initiative of
the Constantine the Philosopher University
under grant no. I-06-399-01, and by
the  Air Force Office of Scientific Research
under contract  FA9550-10-1-0064.


\end{document}